# Lifetimes of spherically symmetric closed universes

Gregory A. Burnett[*]

*Department of Mathematics, North Carolina State University, Raleigh, NC 27695-8205*

(6 September 1994)

It is proven that any spherically symmetric spacetime that possesses a compact Cauchy surface $\Sigma$ and that satisfies the dominant-energy and non-negative-pressures conditions must have a finite lifetime in the sense that all timelike curves in such a spacetime must have a length no greater than $10 \max_\Sigma (2m)$, where $m$ is the mass associated with the spheres of symmetry. This result gives a complete resolution, in the spherically symmetric case, of one version of the closed-universe recollapse conjecture (though it is likely that a slightly better bound can be established). This bound has the desirable properties of being computable from the (spherically symmetric) initial data for the spacetime and having a very simple form. In fact, its form is the same as was established, using a different method, for the spherically symmetric massless scalar field spacetimes, thereby proving a conjecture offered in that work. Prospects for generalizing these results beyond the spherically symmetric case are discussed.

04.20.Dw, 04.20.Cv, 98.80.Hw

---

[*]Present Address: Department of Physics, University of Florida, Gainesville, FL 32611

# I. INTRODUCTION

An open question in classical general relativity is whether a spacetime with $S^3$ or $S^1 \times S^2$ Cauchy surfaces and "ordinary" matter can exist eternally in the sense that it admits arbitrarily long timelike curves. Eternal spacetimes with non-compact Cauchy surfaces are abundant: e.g., the asymptotically flat spacetimes such as Minkowski and Kerr, and the forever expanding open cosmological models such as the $k = 0$ and $k = -1$ Robertson-Walker spacetimes (with, e.g., dust or radiation as matter). Further, eternal spacetimes $(M, g_{ab})$ with compact Cauchy surfaces are easily constructed from any closed, orientable, three-manifold $\Sigma$ that admits a flat metric $h_{ab}$ thereon, such as the three-torus $S^1 \times S^1 \times S^1$. Simply take $M = \mathbb{R} \times \Sigma$ and $g_{ab} = -(dt)_a(dt)_b + a^2(t)h_{ab}$. Then, for example, with $a(t) = 1$, the spacetime is flat and static, while with $a(t) = t^{2/3}$ ($t > 0$) the spacetime is forever expanding (to the future) and has the stress-energy tensor of dust (a perfect fluid with zero pressure). However, the only known eternal spacetimes with $S^3$ or $S^1 \times S^2$ Cauchy surfaces have "peculiar" matter content in that they have negative pressures. For example, while we can construct Robertson-Walker spacetimes that expand forever and satisfy the traditional energy conditions such as the dominant-energy and timelike-convergence conditions (such a choice is $a(t) = t$, where $a(t)$ is the radius of the universe), it is impossible to make this choice so that the spacetime is both eternal and has non-negative pressures [1]. Similarly, the spatially homogeneous spacetimes with $S^3$ or $S^1 \times S^2$ Cauchy surfaces (the Bianchi IX and Kantowski-Sachs spacetimes, respectively) also have finite lifetimes if the dominant-energy and non-negative-pressures conditions are satisfied [2,3].

While there are no *known* eternal spacetimes with $S^3$ or $S^1 \times S^2$ Cauchy surfaces and "ordinary" matter, are there any at all? The *closed-universe recollapse conjecture* asserts there are none [4–7,3,8,9].

A strong form of this conjecture asserts that spacetimes with $S^3$ or $S^1 \times S^2$ Cauchy surfaces and "ordinary" matter expand from an initial singularity to a maximal hypersurface and then recollapse to a final singularity [4–7]. It is this form the conjecture that leads us to restrict our consideration to $S^3$ and $S^1 \times S^2$ the Cauchy surface topologies. For should a maximal hypersurface exist, then by the scalar constraint equation of general relativity and the dominant-energy (or merely non-negative-energy) condition, the scalar curvature associated with the metric induced on the maximal hypersurface must be non-negative. However, very few three-manifolds admit metrics with non-negative scalar curvature [10]. Those that do are $S^3$, $S^1 \times S^2$, those that can be constructed from these by making connected summations and certain identifications, and the three-manifolds admitting flat metrics [11]. The latter are eliminated from consideration by arguing either: that only the flat static spacetimes with such a spatial topology actually admit a maximal Cauchy surface, and therefore, either a maximal hypersurface does not exist or it neither expands nor recollapses; or that eternal spacetimes with such a spatial topology are easily constructed as was done above. So, while in addition to $S^3$ and $S^1 \times S^2$, we can include in our conjecture such manifolds as $(S^1 \times S^2) \# (S^1 \times S^2)$ (where $A \# B$ denotes the connected sum of two manifolds $A$ and $B$ [12]) or $\mathbb{R}P^3$ ($S^3$ with antipodal points identified), we have not done so here for simplicity's sake.

A weaker form of this conjecture merely asserts that all spacetimes with $S^3$ or $S^1 \times S^2$ Cauchy surfaces and "ordinary" matter have finite lifetimes in the sense that there will exist a finite upper bound to the lengths of all timelike curves therein [7,3,8,9]. One precise version of this conjecture is the following.

**Conjecture:** There exists an upper bound to the lengths of timelike curves in any spacetime that possesses $S^3$ or $S^1 \times S^2$ Cauchy surfaces and that satisfies the dominant-energy and non-negative-pressures conditions.

Here, the dominant-energy condition is the demand that $G_{ab}t^a u^b \geq 0$ for all future-directed $t^a$ and $u^b$, and the non-negative-pressures condition is the demand that $G_{ab}x^a x^b \geq 0$ for all spacelike $x^a$. Placing our conditions on the Einstein tensor directly, rather than on the stress-energy tensor, allows us to make arguments independent of the exact theory of gravity being studied (as long as it is a metric theory). So, in the case of Einstein's theory, $G_{ab} = 8\pi T_{ab}$, whether these conditions are satisfied depends entirely on whether these conditions are satisfied by the total stress-energy tensor of the matter fields.

It has been known for nearly a quarter of a century now that closed universes with "ordinary" matter are generically singular. Hawking and Penrose's 1970 theorem states that a spacetime with compact Cauchy surfaces satisfying the timelike convergence condition ($R_{ab}t^a t^b \geq 0$ for all timelike $t^a$) and a genericity condition cannot be both timelike and null geodesically complete [13]. At least one inextendible timelike or null geodesic is incomplete to the future or the past. Unfortunately, this theorem tells us neither whether just a few, most, or all causal geodesic are incomplete nor whether the singular behavior occurs to the future or the past. The closed-universe recollapse conjecture promises that, with a further restriction on the Cauchy surface topology and matter content, all timelike geodesics will be incomplete to both the future and the past.

We study the above conjecture for the spherically symmetric spacetimes and our main result is summarized by the following theorem.

**Theorem 1.** The length of any timelike curve in a spherically symmetric spacetime that possesses a compact Cauchy surface $\Sigma$ and that satisfies the dominant-energy and non-negative-pressures conditions must have a length no greater than $10 \max_\Sigma (2m)$ where $m$ is the mass associated with the spheres of symmetry.

While it is true that the spherically symmetric spacetimes are special in the sense that they do not explore the "full degrees of freedom" available to the gravitational field, e.g., the high-degree of symmetry prevents the existence of gravitational waves (at least in vacuum regions), they do offer a platform on which the more general conjecture can begun to be attacked. That is, if the conjecture above is true, proving it for this limited class of spacetimes should give some insight into how the more general case should be approached. This will be discussed further in Sec. IV. If the conjecture is false, the above theorem shows that counterexamples will not be found among the spherically symmetric spacetimes.

The weak version of the closed-universe recollapse conjecture has previously been studied for the spherically symmetric spacetimes and a number of partial results obtained. Throughout these investigations, the strategy has been to construct arguments in terms of the scalar fields $r$, giving the size of the two-spheres of symmetry, and $m$, a measure of the total amount of "mass" associated with a given sphere of symmetry (a sort of quasilocal mass). The two most important facts about $r$ and $m$ is summarized by the following theorem. (For a proof, see Refs. [3,8]. Note that although the theorems as stated in these references require that the Cauchy surface be spherically symmetric, the following relaxes that requirement.)

**Theorem 2.** For any spherically symmetric spacetime that possesses a compact Cauchy surface $\Sigma$ and that satisfies the dominant-energy and radial-non-negative-pressure conditions,

$$r \leq \max_{\Sigma}(2m), \tag{1.1a}$$

$$2m \geq \min_{\Sigma}(r). \tag{1.1b}$$

That is, $r$ is everywhere bounded above by the maximum of $2m$ on any Cauchy surface, while $m$ is everywhere bounded below by the minimum of $r$ on any Cauchy surface. Furthermore, by the non-negative-pressures condition, along any radial timelike geodesic we have

$$\frac{d^2 r}{dt^2} \leq -\frac{m}{r^2}, \tag{1.2}$$

where $t$ is the proper time along the geodesic.

The conjecture was first resolved for the spherically symmetric spacetimes in the case where the Cauchy surfaces have the topology $S^1 \times S^2$ [3]. Using Eq. (1.2), together with the above bounds on $r$ and $m$, it was shown that the length of any timelike curve is bounded by the expression

$$\left(\pi \sqrt{\frac{\max_\Sigma(2m)}{\min_\Sigma(r)}}\right) \max_\Sigma(2m). \tag{1.3}$$

Whether the new bound given by theorem 1 is better (smaller) or worse (larger) than the bound above depends on the magnitude of the term in parentheses. This quantity has a lower bound of $\pi$ and no upper bound.

Therefore, while the bound given by theorem 1 can be somewhat worse than this one, it can be much better.

Unfortunately, the method that had worked so well for the $S^1 \times S^2$ case, fails for the $S^3$ case. The difference between the two cases is simply that $m$ is merely non-negative in the $S^3$ case and not bounded away from zero as in the $S^1 \times S^2$ case. Clearly, a new approach was needed. Therefore, in an attempt to gain insight into the problem, the conjecture was investigated in a few special cases.

The conjecture was next studied for the dust-filled spherically symmetric spacetimes (the Tolman or Tolman-Bondi spacetimes) possessing $S^3$ Cauchy surfaces. Using a number of properties particular to these spacetimes (e.g., the existence of a geodetic vector field and a preferred globally defined time function), again an upper bound on the lengths of timelike curves was established. Unfortunately, the bound constructed was exceedingly complicated, and worse, the method was just too specialized to the spacetimes being considered which obscured and gave little hope for any generalization.

Next, the conjecture was studied for a class of spherically symmetric spacetimes that included the spherically symmetric massless scalar field spacetimes. Using the fact that $D^a D_a r$ is non-negative for these spacetimes, an application of Stokes's theorem and the use of Eq. (1.2) made the establishment of a bound quite easy. [The derivative operator $D_a$ here is not the one associated with the metric $g_{ab}$. Its definition can be found in Sec. II.] Furthermore, the simplicity of the bound constructed ($6 \max_\Sigma(2m)$) led to the conjecture that a similar bound held more generally. Theorem 1 proves this conjecture true. Yet, while the method used in this case takes very little advantage of the specialness of the spacetimes being considered, a generalization was neither apparent nor found. (Though, is still seems likely that one exists.)

Here, we use yet another method having the great advantages of being applicable to the general spherically symmetric spacetime and producing the simple global upper bound on the lengths of timelike curves given by theorem 1. The ideas that motivated this method can be summarized as follows.

First, as is argued in Sec. III, it is necessary and sufficient that the bound given by theorem 1 hold on the distance $d(\mathcal{S}_1, \mathcal{S}_2)$ between any two spheres of symmetry $\mathcal{S}_1$ and $\mathcal{S}_2$ with $\mathcal{S}_2 \subset I^+(\mathcal{S}_1)$. Connecting these two spheres of symmetry is a (non-unique) causal curve $\mu$ that achieves the maximal length $d(\mathcal{S}_1, \mathcal{S}_2)$. This curve is timelike, geodetic, with $r$ strictly positive thereon. In the $S^1 \times S^2$ case, the length of such a curve is easily bounded by using Eq. (1.2) together with the positive lower bound on $m$ given by theorem 2. However, as has been noted, in the $S^3$ case this method fails.

The key new idea in bounding the length of $\mu$ was to search for another timelike curve $\alpha$ (possibly non-geodetic) connecting $\mathcal{S}_1$ to $\mathcal{S}_2$ and bound the length of $\mu$ in terms of this new curve. But what to use for $\alpha$? What other "natural" curves are there from $\mathcal{S}_1$ to $\mathcal{S}_2$?

One possibility is suggested by constructing the spherically symmetric timelike three-surface $\mathcal{T}$ of maximal three-area that connects $\mathcal{S}_1$ to $\mathcal{S}_2$ and taking $\alpha$ to be a radial timelike curve from $\mathcal{S}_1$ to $\mathcal{S}_2$ that is tangent to $\mathcal{T}$. (See Fig. 1.) The three-area of $\mathcal{T}$ is then simply $\int_\alpha (4\pi r^2)\, d\tau$. (This type of integral is discussed in Sec. III A.) Since $\alpha$ maximizes the three-area, we have $\int_\mu (4\pi r^2)\, d\tau \leq \int_\alpha (4\pi r^2)\, d\tau$. From this, we have a bound on the length of $\mu$ in terms of the length of $\alpha$ given by

$$\text{(length of } \mu) \leq \frac{\langle r^2 \rangle_\alpha}{\langle r^2 \rangle_\mu} \text{(length of } \alpha), \tag{1.4}$$

where $\langle f \rangle_\sigma$ denotes the average of a quantity $f$ over a timelike curve $\sigma$. Defining $K = J^+(\mathcal{S}_1) \cap J^-(\mathcal{S}_2)$, the average of $r^2$ over $\alpha$ is bounded above by $\max_K (r)^2$ (which in turn is bounded above by the square of the bound for $r$ given by theorem 2). Now, for Eq. (1.4) to be at all useful, we need a bound on the length of $\alpha$. Remarkably, we can show that the length of $\alpha$ is bounded by $\pi \max_K(r)$ which in turn is bounded as $r$ is bounded.

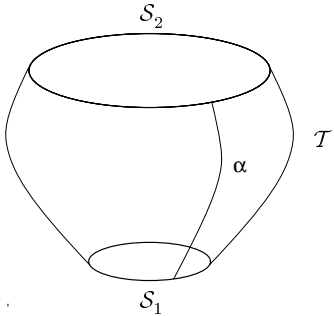

FIG. 1. A figurative representation (one spatial-dimension suppressed so that spheres are represented by circles) of the timelike three-surface $\mathcal{T}$ of maximum three-area that is spherically symmetric and connects the sphere of symmetry $\mathcal{S}_1$ to the sphere of symmetry $\mathcal{S}_2$. The timelike curve $\alpha$ here is radial and tangent to $\mathcal{T}$.

But, what of the average value of $r^2$ on $\mu$? Using the fact that $m$ is non-negative and Eq. (1.2), it follows that $r$ is concave function of $t$ on $\mu$. Therefore, the minimum of $r$ on $\mu$ occurs at either endpoint of $\mu$, thereby allowing us the place a lower bound on the average value of $r^2$ in terms of $r^2$ at its endpoints.

Putting this all together, we have

$$\text{(length of } \mu) \leq \pi \left( \frac{\max_K(r)}{\min_\mu(r)} \right)^2 \max_K(r). \tag{1.5}$$

However, as $r$ at the endpoints of $\mu$ can be arbitrarily small, so can $\min_\mu(r)$. It therefore appears that this bound is useless. Again using the fact that $r$ is a concave function of $t$ on $\mu$, we have by lemma A2 in Appendix A that $\langle r^2 \rangle_\mu \geq \frac{1}{3}(\max_\mu(r))^2$. We now have

$$\text{(length of } \mu) \leq 3\pi \left( \frac{\max_K(r)}{\max_\mu(r)} \right)^2 \max_K(r). \tag{1.6}$$

Since $r$ is bounded by theorem 2, we have the encouraging result that as long as $r$ is *somewhere* "large" on $\mu$, the length of $\mu$ must be "small". Or, put another way, the only way $\mu$ can have a long length is if $r$ is everywhere small on the curve, e.g., if $\mu$ is everywhere near either curve $\gamma_n$ or $\gamma_s$ where $r = 0$. Although this is much better, this still far from our desired result. After all, just as $\min_\mu(r)$ can be arbitrarily small, so can $\max_\mu(r)$.

To obtain an upper bound on the length of $\mu$ that is independent of $\mu$, we use Eq. (1.5) and a bit of trickery. Very briefly, by splitting the curve $\mu$ into three parts, we perform a construction that either "succeeds" which allows us to construct another geodesic on which $r$ is "large" at its endpoints and thereby bounding its length and hence the length of $\mu$, or it "fails" from which a "small region" on which $r$ is "small" is constructed. If it "fails", we repeat the construction on the "small region". Eventually, such a sequence of constructions succeeds and, by the way these regions are constructed, thereby allows us to again bound the length of $\mu$.

Therefore, using the curve $\alpha$ that maximizes the integral $\int_\sigma (4\pi r^2)\, d\tau$ over the set of continuous causal curves that connect $\mathcal{S}_1$ to $\mathcal{S}_2$, we can achieve a bound on the length of $\mu$. However, it will turn out that we eventually get a better bound on the length of $\mu$ by using $\sqrt{r}$ in place of $4\pi r^2$. Therefore, the argument presented in Sec. III constructs $\alpha$ using $\sqrt{r}$ although a geometric interpretation of the integral $\int_\alpha \sqrt{r}\, d\tau$ is not apparent.

In Sec. II, the basics of the spherically symmetric spacetimes are briefly reviewed. In Sec. III, the full details of the proof of theorem 1 are given. Lastly, in Sec. IV, the results presented here and prospects for attacking the non-spherically symmetric case are discussed.

Our conventions are those of Ref. [14]. In particular, metrics are such that timelike vectors have negative norm and the Riemann and Ricci tensors are defined by $2\nabla_{[a}\nabla_{b]}\omega_c = R_{abc}{}^d \omega_d$ and $R_{ab} = R_{amb}{}^m$ respectively. All metrics are taken to be $C^2$. Our units are such that $G = c = 1$.

## II. REVIEW

In this section, the basic features of the spherically symmetric spacetimes needed here are briefly reviewed. For a more complete presentation, see Refs. [3] and [8].

A spacetime $(M, g_{ab})$ is said to be spherically symmetric if it admits a group $G \approx SO(3)$ of isometries, acting effectively on $M$, each of whose orbits is either a two-sphere or a point. Denote the orbit of a point $p$ by $\mathcal{S}_p$. The value of the non-negative scalar field $r$ at each $p \in M$ is defined so that $4\pi r^2$ is the area of $\mathcal{S}_p$. So, in particular, $r(p) = 0$ if $\mathcal{S}_p = p$, while $r(p) > 0$ if $\mathcal{S}_p$ is a two-sphere. Furthermore, we shall say that $\mathcal{S}$ is a sphere of symmetry if $\mathcal{S} = \mathcal{S}_p$ for some $p \in M$ and $\mathcal{S}$ is a two-sphere.

Where $r > 0$, we decompose the metric $g_{ab}$ into the sum $g_{ab} = h_{ab} + q_{ab}$, where $q^a{}_b$ is the projection operator onto the tangent space of each sphere of symmetry and $h^a{}_b$ is the projection operator onto the tangent space of each two-surface perpendicular to the spheres of symmetry. Using the fact that there exists a preferred "unit-metric" $\Omega^{ab}$ on each sphere of symmetry, we have $q_{ab} = r^2 \Omega_{ab}$ (where $\Omega^{am}\Omega_{mb} = q^a{}_b$ and $\Omega_{ab} = q^m{}_a q^n{}_b \Omega_{mn}$). This gives us the final decomposition of $g_{ab}$ as

$$g_{ab} = h_{ab} + r^2 \Omega_{ab}. \tag{2.1}$$

In addition to the derivative operator $\nabla_a$ associated with metric $g_{ab}$, we have the derivative operator $D_a$ associated with the (unphysical) metric $h_{ab} + \Omega_{ab}$. For our purposes, only a single property of $D_a$ need be remembered. If $v^b$ is a spherically symmetric vector field (and therefore radial), then $D_a v_b = h^c{}_a h^d{}_b \nabla_c v_d$. Therefore, if desired, $D_a D_b r$ can be thought of as the "purely radial part" of $\nabla_a \nabla_b r$. It follows that along a radial curve with tangent vector $u^a$, we have $u^a D_a u^b = u^a \nabla_a u^b$. (Of course, since $D_a$ and $\nabla_a$ are both derivative operators, $D_a f = \nabla_a f$ for all scalar fields $f$.)

For the spherically symmetric spacetimes, the mass $m$ associated with the spheres of symmetry is defined by

$$2m = r(1 - D_m r D^m r). \tag{2.2}$$

Lastly, defining $\epsilon^{ab}$ to be either of the two antisymmetric tensor fields such that $\epsilon^{ab}\epsilon^{cd} = -2h^{a[c}h^{d]b}$, and denoting the "radial part" of the Einstein tensor $G^{ab}$ by $\tau^{ab}$ (i.e., $\tau^{ab} = h^a{}_m h^b{}_n G^{mn}$) we have

$$D_a D_b r = \frac{m}{r^2} h_{ab} - \frac{r}{2} \tau^{mn} \epsilon_{ma} \epsilon_{nb}. \tag{2.3}$$

From this, Eq. (1.2) follows by contracting this equation with the unit-tangent vector $t^a$ to the radial geodesic and using the non-negative-pressures condition.

## III. THE ARGUMENT

Fix a spacetime $(M, g_{ab})$ satisfying the conditions of theorem 1 and a Cauchy surface $\Sigma$ therein. Fix $p, q \in M$ with $q \in I^+(p)$. Our goal is to establish the inequality

$$d(p,q) < 10 \max_\Sigma (2m), \tag{3.1}$$

where $d(p, q)$ is the distance function defined as the least upper bound to the length of all continuous causal curves connecting $p$ to $q$ if $q \in J^+(p)$, and zero otherwise [13]. We establish Eq. (3.1) by showing that for any two spheres of symmetry $\mathcal{S}_1$ and $\mathcal{S}_2$ with $\mathcal{S}_2 \subset I^+(\mathcal{S}_1)$ that

$$d(\mathcal{S}_1, \mathcal{S}_2) < 10 \max_\Sigma (2m), \tag{3.2}$$

where $d(\mathcal{P}, \mathcal{Q})$ is defined for subsets $\mathcal{P}$ and $\mathcal{Q}$ of $M$ as the least upper bound of $d(p, q)$ over all $p \in \mathcal{P}$ and $q \in \mathcal{Q}$. That Eq. (3.1) follows from this can be seen as follows. For any $p$ and $q$, we can always find $p' \in I^-(p)$ with $r(p') > 0$ and $q' \in I^+(q)$ with $r(q') > 0$. Then, since any timelike curve from $p$ to $q$ can be extended to a longer timelike curve from $p'$ to $q'$ we have $d(p,q) < d(p', q')$. Furthermore, since $p'$ and $q'$ are subsets of the spheres of symmetry $\mathcal{S}_{p'}$ and $\mathcal{S}_{q'}$ respectively, it follows that $d(p', q') \leq d(\mathcal{S}_{p'}, \mathcal{S}_{q'})$. Therfore, if Eq. (3.2) holds for all spheres of symmetry, in particular $\mathcal{S}_{p'}$ and $\mathcal{S}_{q'}$, Eq. (3.1) follows.

We divide the task of establishing Eq. (3.2) into three parts. In Sec. III A, the integration of scalar fields along causal curves is reviewed and a few characteristics of causal curves that maximize such an integral are pointed out. This construction is then used in Sec. III B where a weak inequality on $d(\mathcal{S}_1, \mathcal{S}_2)$ is established. In Sec. III C, a construction is presented that uses this inequality to establish Eq. (3.2) and thereby theorem 1.

### A. Causal curves maximizing scalar integrals

Given a scalar field $f$ on $M$ and a (differentiable) causal curve $\sigma$ from $p_1$ to $p_2$, the integral of $f$ along $\sigma$, which we denote by $\int_\sigma f \, d\tau$, is defined as follows. Parameterize $\sigma$ by $t$ so that $\sigma(t_1) = p_1$ and $\sigma(t_2) = p_2$ and denote the associated tangent vector to $\sigma$ by $t^a$. Then, we set

$$\int_\sigma f \, d\tau = \int_{t_1}^{t_2} f(\sigma(t)) \sqrt{-g_{ab} t^a t^b} \, dt. \tag{3.3}$$

Although this integral does depend on the path $\sigma$ chosen to connect $p_1$ to $p_2$, it is independent of the choice of parameterization. Further, if $\sigma$ is everywhere null, then the above integral is zero, while if $\sigma$ is a timelike curve, then taking $t$ so that $t^a$ is unit-timelike, we have

$$\int_\sigma f \, d\tau = \int_{t_1}^{t_2} f(\sigma(t)) \, dt. \tag{3.4}$$

As a simple example, taking $f = 1$, then $\int_\sigma d\tau$ is simply the length of the curve $\sigma$.

Let $\alpha$ be any causal curve that maximizes the integral $\int_\sigma f \, d\tau$ over the set of all (continuous) causal curves $\sigma$ from $p_1$ to $p_2$ (should such a curve exist). Then, by the construction of the curve $\alpha$, for any other curve $\gamma$ from $p_1$ to $p_2$, we have

$$\int_\gamma f \, d\tau \leq \int_\alpha f \, d\tau. \tag{3.5}$$

Defining $\langle f \rangle_\sigma$, the average value of $f$ over $\sigma$, for $\sigma$ having non-zero length by

$$\langle f \rangle_\sigma = \frac{1}{(\text{length of } \sigma)} \int_\sigma f \, d\tau, \tag{3.6}$$

from Eq. (3.5) we have the two (equivalent) inequalities on the length of $\gamma$ that will serve a fundamental role in the proof of theorem 1 (when $f$ is chosen appropriately):

$$\text{(length of } \gamma) \leq \frac{1}{\langle f \rangle_\gamma} \int_\alpha f \, d\tau, \tag{3.7a}$$

$$\text{(length of } \gamma) \leq \frac{\langle f \rangle_\alpha}{\langle f \rangle_\gamma} \text{(length of } \alpha), \tag{3.7b}$$

where it has been assumed that $\langle f \rangle_\gamma > 0$ (as will be the case).

For $f$ non-negative, and $(M, g_{ab})$ globally hyperbolic, there does exist a curve $\alpha$ from $p_1$ to $p_2$ that maximizes the integral $\int_\sigma f \, d\tau$ over the set of continuous causal curves from $p_1$ to $p_2$. (The proof is merely a slight modification of the proof of the existence of a maximal length causal curve connecting two points in a globally hyperbolic spacetime. See, e.g., Ref. [14].) For $f$ positive everywhere (or at least on $J^+(p_1) \cap J^-(p_2)$), then by its definition, $\int_\sigma f \, d\tau$ is simply the length of the curve $\sigma$ measured using the metric $\tilde{g}_{ab} = f^2 g_{ab}$. Therefore, in this case, the curve that maximizes the integral $\int_\sigma f \, d\tau$ is simply a geodesic in the spacetime $(M, \tilde{g}_{ab})$. Hence, with $p_2 \in I^+(p_1)$, this curve must be timelike. Carrying out the variation of the integral $\int_\sigma f \, d\tau$ over a one-parameter family of differentiable timelike curves, we find that a necessary condition for the curve $\alpha$ to maximize the integral is that

$$u^b \nabla_b (f u^a) + \nabla^a f = 0, \tag{3.8}$$

where $u^a$ is a unit-tangent vector to the curve $\alpha$. Alternatively, we could have arrived at this equation by writing down the geodesic equation for $\tilde{g}_{ab}$ and then reexpressing it in terms of $u^a$ and the connection associated with $g_{ab}$.

When $f$ is merely non-negative, the character of a causal curve that maximizes $\int_\sigma f \, d\tau$ is not so clear. Must the curve be differentiable? Must it be timelike everywhere? For instance, when $f = 0$, any causal curves maximizes this integral; each has total integral zero. Therefore, a curve maximizing the integral in this case need neither be differentiable nor everywhere timelike. This indicates that the study of curves maximizing $\int_\sigma f \, d\tau$ when $f$ can be zero requires some care. However, it can be shown that when $f$ is positive on the curve $\alpha$ that maximizes $\int_\sigma f \, d\tau$, then again, if $p_2 \in I^+(p_1)$, the curve $\alpha$ is everywhere timelike and satisfies Eq. (3.8).

Lastly, for $f = g(r)$, where $g$ is a continuous function that is positive (and differentiable) for $r$ positive, a curve $\alpha$ that maximizes the integral $\int_\sigma f \, d\tau$ over all continuous causal curves from a sphere of symmetry $\mathcal{S}_1$ to another sphere of symmetry $\mathcal{S}_2 \subset I^+(\mathcal{S}_1)$ is a radial curve with $r$ positive everywhere thereon (i.e., just as maximal geodesics connecting $\mathcal{S}_1$ to $\mathcal{S}_2$ don't intersect the curves where $r$ is zero, neither does $\alpha$) and therefore is timelike and also obeys Eq. (3.8).

### B. A key lemma

In this section, we establish the following lemma which places an upper bound on the distance between two spheres of symmetry. With a further construction presented in the next section, this allows us to complete the proof of theorem 1.

**Lemma 1.** Fix two spheres of symmetry $\mathcal{S}_1$ and $\mathcal{S}_2$ with $\mathcal{S}_2 \subset I^+(\mathcal{S}_1)$ and set $K = J^+(\mathcal{S}_1) \cap J^-(\mathcal{S}_2)$. Then, for any $\lambda > 0$ such that

$$r(\mathcal{S}_1) \geq \lambda \max_K (r), \tag{3.9a}$$

$$r(\mathcal{S}_2) \geq \lambda \max_K (r), \tag{3.9b}$$

we have

$$d(\mathcal{S}_1, \mathcal{S}_2) \leq \left( \frac{4}{3} \frac{\sqrt{1-\lambda}(2+\lambda)}{\sqrt{\lambda}} \right) \max_K (r). \tag{3.10}$$

*Proof.* Let $\gamma$ be a causal curve of maximal length connecting $\mathcal{S}_1$ to $\mathcal{S}_2$. This curve has length $d(\mathcal{S}_1, \mathcal{S}_2)$, is radial, timelike, geodetic, and since $\gamma$ is a maximal curve with $r$ positive at its endpoints, $r$ is positive on all of $\gamma$ [9]. To prove Eq. (3.10), we establish this bound on the length of $\gamma$ using Eq. (3.7a) with $f = \sqrt{r}$.

Since $m$ is non-negative, by Eq. (1.2), $r$ is a concave function of $t$ on the radial geodesic $\gamma$ and, therefore, the minimum of $r$ on $\gamma$ must occur at either endpoint of $\gamma$ (i.e., on $\mathcal{S}_1$ or $\mathcal{S}_2$). So, by Eq. (3.9), we have

$$\langle \sqrt{r} \rangle_\gamma \geq \sqrt{\min_\gamma (r)} \geq \sqrt{\lambda} \sqrt{\max_K (r)}. \tag{3.11}$$

Let $\alpha$ be a causal curve that maximizes $\int_\sigma \sqrt{r} \, d\tau$ over the set of all continuous causal curves connecting $\mathcal{S}_1$ to $\mathcal{S}_2$. This curve is radial, timelike, with $r$ strictly positive thereon. Although $\alpha$ is not geodetic, using Eq. (3.8) with $f = \sqrt{r}$ and the fact that $\alpha$ is a radial curve, we have

$$u^b D_b u^a = -\frac{1}{2r} \left( h^{ab} + u^a u^b \right) D_b r, \tag{3.12}$$

where $u^a$ is a unit-tangent vector to $\alpha$. From this, the behavior of $r$ on $\alpha$ is restricted by the inequality

$$2r\ddot{r} + \dot{r}^2 + 1 \leq 0, \tag{3.13}$$

where a dot denotes a derivative with respect to the proper time along $\alpha$, i.e., $\dot{r} = u^a D_a r$. To see this, consider

$$\begin{aligned}
\ddot{r} &= u^a D_a (u^b D_b r) \\
&= (u^a D_a u^b) D_b r + u^a u^b D_a D_b r \\
&= -\frac{1}{2r} \left( h^{ab} + u^a u^b \right) (D_a r)(D_b r) \\
&\quad + u^a u^b \left( \frac{m}{r^2} h_{ab} - \frac{r}{2} \tau^{mn} \epsilon_{ma} \epsilon_{nb} \right) \\
&\leq -\frac{1}{2r} \left( \left( 1 - \frac{2m}{r} \right) + \dot{r}^2 \right) - \frac{m}{r^2} \\
&\leq -\frac{1}{2r} (1 + \dot{r}^2). \tag{3.14}
\end{aligned}$$

The third step follows from equation of motion for $\alpha$ given by Eq. (3.12) and the expression for the second derivative of $r$ given by Eq. (2.3). The fourth step follows from the definition of $m$, the fact that $u^a$ is unit-timelike, and the non-negative-pressures condition.

An analysis of Eq. (3.13), given in Appendix A, shows that the maximum time that $r$ can remain positive and bounded by $\max_K(r)$ on $\alpha$ (i.e., $0 < r \leq \max_K(r)$) is $\pi \max_K(r)$. Furthermore, by lemma A1, we have

$$\int_\alpha \sqrt{r}\, d\tau \leq \frac{4}{3}\sqrt{1-\lambda}(2+\lambda)\left(\max_K(r)\right)^{3/2}. \quad (3.15)$$

Therefore, Eq. (3.7a) with our lower bound on $\langle\sqrt{r}\rangle_\gamma$ given by Eq. (3.11) and upper bound on $\int_\alpha \sqrt{r}\, d\tau$ given by Eq. (3.15), establishes Eq. (3.10). $\square$

It should be noted that Eq. (3.15) is merely a finer version of the bound $\int_\alpha \sqrt{r}\, d\tau \leq \pi(\max_K(r))^{3/2}$ which is obtained using the equality $\int_\alpha \sqrt{r}\, d\tau = \langle\sqrt{r}\rangle_\alpha (\text{length of } \alpha)$ and the crude bounds $\langle\sqrt{r}\rangle_\alpha \leq \sqrt{\max_K(r)}$ and $(\text{length of } \alpha) \leq \pi \max_K(r)$.

If $r$ at the endpoints of $\mu$ were of the order of $\max_K(r)$, e.g., so that $\lambda \geq \frac{1}{2}$, then using lemma 1 we would have an upper bound on the length of $\mu$. Things are not so simple. While $r$ at the endpoints of $\mu$ are non-zero, so that the bound given by Eq. (3.10) is finite, $r$ there (and hence $\lambda$) can be arbitrarily small. The $\sqrt{\lambda}$ factor in the denominator of Eq. (3.10) keeps us from establishing a finite upper bound by such a simple argument.

We note that another version of lemma 1 can be obtained by using the lower bound of

$$\langle\sqrt{r}\rangle_\gamma \geq \frac{2}{3}\sqrt{\max_\gamma(r)}, \quad (3.16)$$

which also follows from the fact that $r$ is a concave function of $t$ on $\gamma$ and follows from lemma A2 in Appendix A.

Before going on to apply this lemma to complete the proof of theorem 1, we pause to consider Eq. (3.13). This equation possesses a number of interesting features. First, in vacuum the inequality becomes an equality. This feature distinguishes the choice $f = (\text{constant})\sqrt{r}$ from all others. (See Appendix B.) Second, this inequality is precisely the same one that arises in the study of the Kantowski-Sachs spacetimes for $r(t)$ (again, giving the size of spheres of symmetry as a function of time), and also in the study of the $k = +1$ Robertson-Walker spacetimes for $a(t)$ (giving the size of the universe as a function of time). This is no accident. For consider a Kantowski-Sachs spacetime coordinated so that

$$g_{ab} = -(dt)_a(dt)_b + e^{2\beta(t)}(d\chi)_a(d\chi)_b + r^2(t)\Omega_{ab}, \quad (3.17)$$

where $0 \leq \chi \leq \pi$ and $\chi = 0$ and $\chi = \pi$ are to be identified (as the spatial topology is $S^1 \times S^2$). Consider two spheres of symmetry $\mathcal{S}_1$ at $t = t_1$ and $\chi = \pi/2$ and $\mathcal{S}_2$ at $t = t_2$ and $\chi = \pi/2$. Since $r$ is constant on the surfaces of spatial homogeneity (i.e, surfaces of constant $t$), the curve $\alpha$ that maximizes the integral $\int_\sigma \sqrt{r}\, d\tau$ over all continuous casual curves from $\mathcal{S}_1$ to $\mathcal{S}_2$ is simply a radial geodesic from $\mathcal{S}_1$ to $\mathcal{S}_2$ in the surface $\chi = \pi/2$ and therefore normal to surfaces of homogeneity. In other words, in this case, $\alpha$ coincides with an integral curve of the geodesic flow normal to the surfaces of homogeneity. As the evolution equation for $r$ can be produced by calculating how $r$ changes along one of these integral curves (e.g., using Eq. (2.3), it is no surprise that Eq. (3.13) reproduces the equation giving the evolution of $r$ for these spacetimes. (Note also that this equation is simply Eq. (1.2) rewritten using the fact that for these spacetimes $2m = r(1 + \dot{r}^2)$.) Likewise, consider a Robertson-Walker spacetime coordinated so that

$$g_{ab} = -(dt)_a(dt)_b + a^2(t)\left((d\chi)_a(d\chi)_b + \sin^2\chi\, \Omega_{ab}\right), \quad (3.18)$$

where $0 \leq \chi \leq \pi$. Here, $r(t,\chi) = a(t)\sin\chi$. Again, consider two spheres just as in the Kantowski-Sachs case. Here, although $r$ is not constant on surfaces of spatial homogeneity, it is maximal on the surface $\chi = \pi/2$. Therefore, the curve $\alpha$ is again radial, in the surface $\chi = \pi/2$, and therefore normal to the surfaces of homogeneity. Therefore, as $\alpha$ again coincides with an integral curve of the geodetic flow normal to the surfaces of homogeneity and since $r(t, \pi/2) = a(t)$, it is again no accident that Eq. (3.13) reproduces the equation giving the evolution of $a$ for these spacetimes.

### C. Final construction

With lemma 1 in hand, we can now complete the proof of theorem 1. For any number $0 < \lambda < \frac{1}{2}$, consider the following construction.

Fix two spheres of symmetry $\mathcal{S}_1$ and $\mathcal{S}_2$ with $\mathcal{S}_2 \subset I^+(\mathcal{S}_1)$ and a causal curve $\mu$ of maximal length connecting $\mathcal{S}_1$ to $\mathcal{S}_2$. This curve has length $d(\mathcal{S}_1, \mathcal{S}_2)$, is radial, timelike, geodetic, and since $\mu$ is a maximal curve with $r$ positive at its endpoints, $r$ is positive on all of $\mu$. Denote the past and future endpoints of $\mu$ by $p$ and $q$ respectively, so $\mathcal{S}_p = \mathcal{S}_1$ and $\mathcal{S}_q = \mathcal{S}_2$. Define $p'$ and $q'$ to be those two points on $\mu$ such that

$$d(p, p') = \lambda d(p, q), \quad (3.19a)$$
$$d(q', q) = \lambda d(p, q). \quad (3.19b)$$

By the maximality of $\mu$, for any two points $a$ and $b$ on $\mu$ (with $b \in I^+(a)$) $d(a, b)$ is the length of the segment of $\mu$ connecting $a$ to $b$. Therefore, $p'$ ($q'$) is simply the point on $\mu$ a distance $\lambda$ times the length of $\mu$ measured from the past (future) endpoint of $\mu$. Further,

$$d(p', q') = (1 - 2\lambda)d(p, q). \quad (3.20)$$

Set

$$K = J^+(\mathcal{S}_p) \cap J^-(\mathcal{S}_q), \quad (3.21a)$$
$$P = J^+(\mathcal{S}_p) \cap J^-(\mathcal{S}_{p'}), \quad (3.21b)$$
$$Q = J^+(\mathcal{S}_{q'}) \cap J^-(\mathcal{S}_q). \quad (3.21c)$$

Suppose that

$$\max_P(r) \geq \lambda \max_K(r), \qquad (3.22a)$$
$$\max_Q(r) \geq \lambda \max_K(r). \qquad (3.22b)$$

Let $a$ and $b$ be two points where $r$ reaches its maximum value on the compact sets $P$ and $Q$ respectively, i.e., so that

$$r(a) = \max_P(r), \qquad (3.23a)$$
$$r(b) = \max_Q(r). \qquad (3.23b)$$

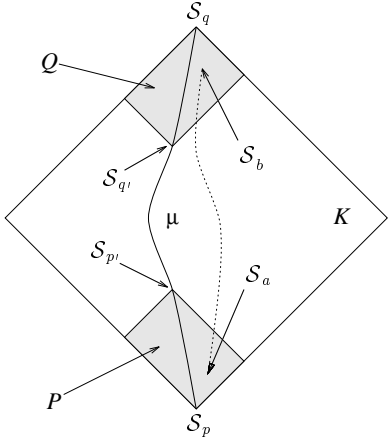

FIG. 2. The curve of maximal length $\mu$ from $\mathcal{S}_p = \mathcal{S}_1$ to $\mathcal{S}_q = \mathcal{S}_2$ is divided into three parts: $p$ to $p'$, $p'$ to $q'$, and $q'$ to $q$. The length of the portion of $\mu$ from $p'$ to $q'$ (being $(1-2\lambda)$(length of $\mu$)) is no longer than the distance from the sphere $\mathcal{S}_a$ (where $r$ is maximal on $P$) to the sphere $\mathcal{S}_b$ (where $r$ is maximal on $Q$). (The dotted curve represents a timelike curve of maximal length from $\mathcal{S}_a$ to $\mathcal{S}_b$.)

Then, we have

$$\begin{aligned} d(\mathcal{S}_1, \mathcal{S}_2) &= \frac{1}{1-2\lambda} d(p', q') \\ &= \frac{1}{1-2\lambda} d(\mathcal{S}_{p'}, \mathcal{S}_{q'}) \\ &\leq \frac{1}{1-2\lambda} d(\mathcal{S}_a, \mathcal{S}_b) \\ &\leq F(\lambda) \max_K(r), \end{aligned} \qquad (3.24)$$

where we have set

$$F(\lambda) = \frac{4}{3} \frac{\sqrt{1-\lambda}(2+\lambda)}{(1-2\lambda)\sqrt{\lambda}}. \qquad (3.25)$$

(See Fig. 2.) The first step follows from Eq. (3.20) and the fact that $d(\mathcal{S}_1, \mathcal{S}_2) = d(p, q)$. The second step follows from the fact that, while in general $d(p', q') \leq d(\mathcal{S}_{p'}, \mathcal{S}_{q'})$, we have an equality as a consequence of the spherical symmetry and the maximality of the length of $\mu$ in connecting $\mathcal{S}_1$ and $\mathcal{S}_2$. The third step follows from the fact that $\mathcal{S}_a$ lies to the causal future of $\mathcal{S}_{q'}$ while $\mathcal{S}_b$ lies to the causal past of $\mathcal{S}_{p'}$ and therefore $d(\mathcal{S}_a, \mathcal{S}_b)$ can be no less than $d(\mathcal{S}_{p'}, \mathcal{S}_{q'})$. The last step follows from lemma 1 that was established in Sec. III B and the fact that $(J^+(\mathcal{S}_a) \cap J^-(\mathcal{S}_b)) \subset K$.

Next, suppose that Eq. (3.22) does not hold. In this case, let $K_1$ and $\mu_1$ denote either region and associated segment, respectively, for which the inequality is violated, i.e., either $P$ and the segment of $\mu$ from $p$ to $p'$ or $Q$ and the segment of $\mu$ from $q'$ to $q$. We now perform exactly the same construction on $\mu_1$ that we performed on $\mu$. However, again the requisite inequalities (the analogs of Eq. (3.22)) may fail. If so, we then construct a segment $\mu_2$ and $K_2$ from $\mu_1$, in the same way we constructed $\mu_1$ and $K_1$ from $\mu$, and continue again.

Eventually, the inequalities analogous to Eq. (3.22) must be satisfied. We can see this as follows. Suppose that it always failed. Then, since

$$\max_{K_{k+1}}(r) < \lambda \max_{K_k}(r), \qquad (3.26)$$

and $\lambda < 1$, $\max_{K_k}(r)$ would become arbitrarily small as $k \to \infty$. By the construction of $K_k$, this would imply the existence of a point on $\mu$ at which $r$ is zero. This is a contradiction as $r$ is positive everywhere on $\mu$, so the construction must eventually succeed.

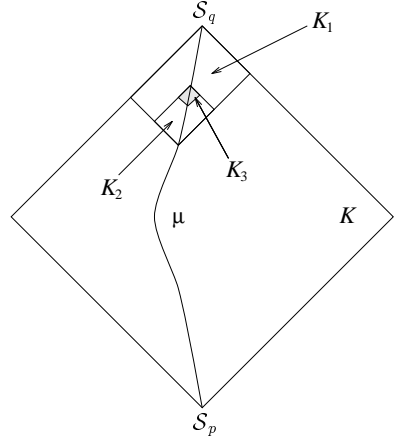

FIG. 3. In the case depicted here, the construction fails for $K$, so $K_1$ is constructed from either region where the requisite inequality fails—in this case the upper region of $K$. Again, the construction fails for $K_1$ so $K_2$ is constructed—this time from the lower region of $K_1$. Again, the construction fails for $K_2$ so $K_3$ is constructed—this time from the upper region of $K_2$. Finally, in this case, the requisite inequalities hold for the region $K_3$.

If the construction above does not succeed with $\mu$, then for some $n \geq 1$ it fails for all $1 \leq k < n$ and succeeds for $k = n$. (In the case depicted in Fig. 3, $n = 3$.) Repeating the argument for the curve $\mu_n$ that was used for $\mu$, we have

$$(\text{length of } \mu_n) \leq F(\lambda) \max_{K_n}(r). \qquad (3.27)$$

Using the facts that

$$d(\mathcal{S}_1, \mathcal{S}_2) = \frac{1}{\lambda^n}(\text{length of } \mu_n), \qquad (3.28)$$

$$\max_{K_n}(r) < \lambda^n \max_K(r). \qquad (3.29)$$

we again find that the length of $\mu$ is bounded by Eq. (3.24) (though with a strict inequality).

It is at this point that we finally use the fact that $r$ is bounded above. By Eq. (3.24) and theorem 2, we have

$$d(\mathcal{S}_1, \mathcal{S}_2) \leq F(\lambda) \max_\Sigma(2m) \qquad (3.30)$$

We are now free to choose $\lambda$ so as to minimize the coefficient $F(\lambda)$. Although the absolute minimum of $F$ on $(0, \frac{1}{2})$ can be found analytically, we obtain a bound that is nearly as good by simply choosing $\lambda = \frac{1}{6}$. We find that $F(\frac{1}{6}) < 9.7$, thereby completing the proof of theorem 1 (with a slightly better bound than was advertised).

## IV. DISCUSSION

Having established an upper bound on the lifetimes of the spherically symmetric spacetimes with compact Cauchy surfaces, we now raise a number of questions concerning this bound and discuss the hopes for generalizing this result beyond the spherically symmetric case.

How good is the bound given by theorem 1? Can it be improved upon? Consider the numerical coefficient 10 in this bound. As has been mentioned, and is further discussed in Appendix B, this number is not optimal and can be reduced slightly (down to 9.31) using a generalization of the methods presented here. Further, it can be shown that even this generalization cannot be optimal. What is the smallest coefficient for which the theorem remains true? For the spherically symmetric massless scalar field spacetimes with compact Cauchy surfaces, the upper bound established has the same form with the smaller coefficient 6 instead of 10 [9]. Further, for the $k = +1$ Robertson-Walker spacetimes and the Kantowski-Sachs spacetimes as well, again a similar bound holds with the coefficient being $\pi$. For these spacetimes this is least possible value for this coefficient as the lifetime of such a spacetime with dust is $\pi \max_\Sigma(2m)$, where here it does not matter which Cauchy surface $\Sigma$ we choose (as $m$ is constant on the flow lines of the dust). Does theorem 1 remain true with the coefficient being $\pi$ rather than 10? Or, does there exist an example showing the necessity of a larger value?

A peculiar property of the bound given by theorem 1 can be seen by evaluating it for various Cauchy surfaces in a $k = +1$ Robertson-Walker spacetime with non-zero pressure, e.g., a radiation dominated model. For any Robertson-Walker spacetime, $m = \frac{4\pi}{3}\rho r^3$, and $r(t, \chi) = a(t) \sin \chi$, and therefore, we have for a Cauchy surface $\Sigma$ that is a surface of homogeneity

$$\max_\Sigma(2m) = \frac{8\pi}{3}\rho(t)a^3(t). \qquad (4.1)$$

For the radiation dominated models, where $P = \frac{1}{3}\rho$, $\rho a^4 = C$ for some constant $C$. By the scalar constraint equation, we find that we can write $C = \frac{3}{8\pi}a_{\max}^2$, which gives

$$\max_\Sigma(2m) = \frac{a_{\max}^2}{a(t)}. \qquad (4.2)$$

From this we see that our bound is least if $\Sigma$ is chosen to be the maximal hypersurface, in which case $\max_\Sigma(2m)$ is just the maximum radius of the universe (a result that holds more generally for the $k = +1$ Robertson-Walker spacetimes) while the bound can be made arbitrarily large by choosing $\Sigma$ when the universe was small, i.e., near the moment of the big bang. (This is a sort of cosmological analog of the mass-inflation phenomena seen in the interior of black holes [15].)

Given initial data for a "young" universe, the upper bound on the lifetime of the universe given by theorem 1 can be much larger than its actual lifetime. Therefore, this bound need not be a good estimate of the actual lifetime. Without going into the details of the matter content, it doesn't seem that we can do any better. After all, if the matter were dust, then in the $k = +1$ Robertson-Walker case, the upper bound would be (within a factor of about 3) the correct lifetime of the spacetime.

Can the conditions on the Einstein tensor (i.e., the energy conditions) in theorem 1 be weakened so that the conclusion of the theorem (or a similar version) remains true? It is clear from the method of proof, that the stated conditions (dominant-energy and non-negative-pressures) need only hold on the "radial part" of the Einstein tensor, i.e., $\tau^{ab}$. This slight relaxation thereby allows us to conclude that the closed spherically symmetric massless scalar field spacetimes have finite lifetimes—a result established previously by a different method [9]. However, can a more significant weakening be attained? In particular, the requirement that the pressures are nowhere negative is very strong. Does a result similar to theorem 1 hold with negative pressures so long as they are not large compared to the energy density? Whether this is the case is unknown.

Now that we know that counterexamples to the version of the closed-universe recollapse conjecture stated in the Introduction are not to be found among the spherically symmetric spacetimes, are any to be found elsewhere? If not, how do we prove a theorem similar to theorem 1 that relaxes the restriction to spherical symmetry?

While at present these questions remain unanswered, the method of proof of theorem 1 does give a few hints as to a possible direction for proving the general case. In the proof of theorem 1, the spheres of symmetry play a central role. For instance, we have $r$ giving their size, $m$ giving a sort of quasilocal mass associated with each, and in lemma 1 we establish an upper bound on the distance between pairs of such spheres. Therefore, it would seem that any attempt at adapting the proof used here to the general case would require a substitute for these spheres. Perhaps any two-spheres would suffice in this role, or perhaps, only certain special two-spheres need be considered. As a part of this, a successful generalization would seemingly require the notion of a quasilocal mass for each sphere. While there are numerous proposals for such masses, which one, if any, is appropriate should suggest itself in the course of a proof. If the proof of the spherically symmetric case is any indication, such a quasilocal mass will be everywhere non-negative (a quasilocal version of the positive energy theorem) and (one would hope) one should be able to bound the sizes of the spheres everywhere in terms of this quasilocal mass on a Cauchy surface (the analog of theorem 2).

However, there seem to be a number of difficulties in adapting the proof here to the general case. Recall that in the Introduction we constructed a curve $\alpha$ by constructing the timelike three-surface $\mathcal{T}$ having maximal three-area over all timelike spherically symmetric three-surfaces connecting two spheres of symmetry. However, if we drop the requirement of spherical symmetry, the construction of $\mathcal{T}$ fails. Given two two-spheres (one to the future of the other), we can construct a sequence of timelike three-surfaces $\mathcal{T}_i$ whose associated sequence of three-areas is unbounded (i.e., there can be no $\mathcal{T}$ with maximal area). So, such an obvious generalization fails. Perhaps a more subtle analog of $\mathcal{T}$ will work (i.e., a surface $\mathcal{T}$ that maximizes not its three-area, but instead an integral involving its extrinsic curvature). However, even if this succeeds, how do we then use $\mathcal{T}$ to bound the length of the curve $\mu$ that maximizes the distance between the two two-spheres?

Therefore, while the method of proof used here in the spherically symmetric case does give a few hints as to how to attack the more general case, at this point it is still unclear as to how to proceed. At best, these hints are subtle; at worst, they are misleading.

### ACKNOWLEDGMENTS

I would like to thank Robert Geroch and Robert Wald for their comments in the course of this work.

### APPENDIX A: TWO LEMMAS

**Lemma A1.** Fix a positive function $r$ on $[t_1, t_2]$ satisfying

$$2r\ddot{r} + \dot{r}^2 + 1 \leq 0, \qquad (A1)$$

where a dot denotes a derivative with respect to $t$. Then for any upper bound $r_U$ of $r$ on $[t_1, t_2]$ ($r \leq r_U$) and $\lambda_1$ and $\lambda_2$ such that

$$r(t_1) \geq \lambda_1 r_U \text{ and } r(t_2) \geq \lambda_2 r_U, \qquad (A2)$$

we have

$$(t_2 - t_1) \leq \pi r_U, \qquad (A3)$$

$$\int_{t_1}^{t_2} \sqrt{r(t)}\, dt \leq (g(\lambda_1) + g(\lambda_2)) r_U^{3/2}, \qquad (A4)$$

where $g(\lambda) = \frac{2}{3}\sqrt{1-\lambda}(2+\lambda)$.

*Proof.* Since $\ddot{r} < 0$, we can divide the interval $[t_1, t_2]$ into two subintervals: that portion on which $\dot{r} \geq 0$; and that portion on which $\dot{r} \leq 0$. Suppose the interval on which $\dot{r} \leq 0$ is non-empty. In this case, there is a $t_0$ ($t_1 \leq t_0 < t_2$) so that this interval is $[t_0, t_2]$.

Multiplying Eq. (A1) by $\dot{r}$ we find that on $[t_0, t_2]$

$$\frac{d}{dt}(r\dot{r}^2 + r) \geq 0. \qquad (A5)$$

Integrating this equation we have

$$r\dot{r}^2 + r \geq r(t_0)(\dot{r}(t_0))^2 + r(t_0), \qquad (A6)$$

and therefore

$$r\dot{r}^2 \geq r(t_0) - r. \qquad (A7)$$

Parameterizing $r$ by $\eta$ through the relationship

$$r(\eta) = r(t_0) \cos^2(\eta/2), \qquad (A8)$$

where the parameter $\eta$ satisfies $0 \leq \eta \leq \eta_2 \leq \pi$, Eq. (A7) becomes

$$\frac{dt}{d\eta} \leq r(t_0) \cos^2(\eta/2). \qquad (A9)$$

Integrating this inequality, we find

$$(t_2 - t_0) \leq \frac{1}{2} r(t_0)(\eta_2 + \sin \eta_2) \leq \frac{\pi}{2} r(t_0) \leq \frac{\pi}{2} r_U. \qquad (A10)$$

Repeating this argument in the case where $\dot{r} \geq 0$, we find that $(t_0 - t_1) \leq \frac{\pi}{2} r_U$ which, with the above, establishes Eq. (A3).

Next, consider

$$\int_{t_0}^{t_2} \sqrt{r(t)}\, dt = \int_0^{\eta_2} \sqrt{r(\eta)} \frac{dt}{d\eta}\, d\eta$$

$$\leq (r(t_0))^{3/2} \int_0^{\eta_2} \cos^3(\eta/2)\, d\eta$$

$$\leq \frac{2}{3}(r(t_0))^{3/2} \sin(\eta_2/2)\left(2 + \cos^2(\eta_2/2)\right)$$

$$\leq \frac{2}{3}(r(t_0))^{3/2} \sqrt{1 - \frac{r(t_2)}{r(t_0)}}\left(2 + \frac{r(t_2)}{r(t_0)}\right)$$

$$\leq g\left(\frac{r(t_2)}{r(t_0)}\right)(r(t_0))^{3/2}$$

$$\leq g(\lambda_2) r_U^{3/2}.$$

In the last step we used the fact that $r(t_0) \leq r_U$ and the facts that $\lambda_2 \leq r(t_2)/r_U \leq r(t_2)/r(t_0)$ and that $g(\lambda)$ is a decreasing function of $\lambda$. Repeating this argument in the case where $\dot{r} \geq 0$, find that $\int_{t_1}^{t_0} \sqrt{r(t)}\, dt \leq g(\lambda_1) r_U^{3/2}$ which, with the above, establish Eq. (A4). □

**Lemma A2.** Fix a concave function $g$ on $[t_1, t_2]$, set $M = \max_{[t_1,t_2]}(g)$, and let $m$ be any number such that $m \leq \min_{[t_1,t_2]}(g) < M$. Then, for any non-decreasing function function $f$ on $[m, M]$, we have $\langle f \circ g \rangle_{[t_1,t_2]} \geq \langle f \rangle_{[m,M]}$. That is,

$$\frac{1}{t_2 - t_1} \int_{t_1}^{t_2} f(g(t))\, dt \geq \frac{1}{M-m} \int_m^M f(x)\, dx. \quad (A11)$$

*Proof.* Denote by $t_0$ the point at which $g$ reaches its maximum value on $[t_1, t_2]$, i.e., so $g(t_0) = M$. Then, since $g$ is concave, we have

$$g\big((1-\lambda)t_1 + \lambda t_0\big) \geq (1-\lambda)g(t_1) + \lambda g(t_0), \quad (A12)$$

for all $0 \leq \lambda \leq 1$. Therefore, since $g(t_1) \geq m$ and $f$ is non-decreasing, we have

$$f\big(g((1-\lambda)t_1 + \lambda t_0)\big) \geq f\big((1-\lambda)m + \lambda M\big), \quad (A13)$$

for all $0 \leq \lambda \leq 1$. Integrating both sides over $\lambda$ over the interval $[0, 1]$ and performing the change of variables $t = (1-\lambda)t_1 + \lambda t_0$ for the left integral and $x = (1-\lambda)m + M$ for the right integral, we find (if $t_0 \neq t_1$)

$$\frac{1}{t_0 - t_1} \int_{t_1}^{t_0} f(g(t))\, dt \geq \frac{1}{M-m} \int_m^M f(x)\, dx. \quad (A14)$$

Repeating this argument for the interval $[t_0, t_2]$, we find (if $t_0 \neq t_2$)

$$\frac{1}{t_2 - t_0} \int_{t_0}^{t_2} f(g(t))\, dt \geq \frac{1}{M-m} \int_m^M f(x)\, dx. \quad (A15)$$

Combing these results, the lemma follows immediately. □

With $g(t) = r(t)$, $m = 0$, and $f(x) = \sqrt{x}$, we find $\langle \sqrt{r} \rangle_{[t_1,t_2]} \geq \frac{2}{3}\sqrt{\max_{[t_1,t_2]}(r)}$. Similarly, with $f(x) = x^2$ we have $\langle r^2 \rangle_{[t_1,t_2]} \geq \frac{1}{3}(\max_{[t_1,t_2]}(r))^2$.

## APPENDIX B: WHY USE $\sqrt{r}$?

In this appendix we explain why $\sqrt{r}$ was used in Sec. III B and not $4\pi r^2$ or some other function $f(r)$.

If we construct $\alpha$ by maximizing the integral $\int_\sigma f(r)$ over all causal curves $\sigma$ connecting two two-spheres of symmetry, then $\alpha$ is radial, and by Eq. (3.8), satisfies the equation

$$u^b D_b u^a = -\frac{f'}{f}(h^{ab} + u^a u^b) D_b r \quad (B1)$$

where $u^a$ is the unit-tangent vector to $\alpha$. Define the non-negative quantity $Q^2$ by setting

$$Q^2 = (h^{ab} + u^a u^b)(D_a r)(D_b r). \quad (B2)$$

Note that $Q$ is the derivative of $r$ perpendicular to $u^a$ and was used in Refs. [8,9]. With this, we can express $2m$ along $\alpha$ as

$$2m = r(1 + \dot{r}^2 - Q^2). \quad (B3)$$

Repeating the argument used in Eq. (3.14), we find

$$\begin{aligned}
\ddot{r} &= u^a D_a(u^b D_b r) \\
&= (u^a D_a u^b) D_b r + u^a u^b D_a D_b r \\
&\leq -\frac{f'}{f} Q^2 - \frac{m}{r^2} \\
&\leq -\frac{1}{2r}(1 + \dot{r}^2) - Q^2\left(\frac{f'}{f} - \frac{1}{2r}\right).
\end{aligned} \quad (B4)$$

Therefore, we recover Eq. (3.13) if we choose $f$ so that

$$\frac{f'}{f} \geq \frac{1}{2r}. \quad (B5)$$

In particular, the inequality in Eq. (3.13) is an equality in vacuum iff the inequality in Eq. (B5) is an equality, i.e. iff $f = (\text{constant})\sqrt{r}$. The solution to Eq. (B5) is $f(r) = g(r)\sqrt{r}$ where $g$ is non-decreasing ($g' \geq 0$).

We now bound the length of a timelike curve $\gamma$ of maximal length connecting the two spheres of symmetry. We have

$$\int_\alpha f(r)\, d\tau = \int_{t_1}^{t_2} g(r(t))\sqrt{r(t)}\, dt, \quad (B6)$$

and the lower bound

$$\langle f(r) \rangle_\gamma \geq f(\min_\gamma(r)) = g(\min_\gamma(r))\sqrt{\min_\gamma(r)}. \quad (B7)$$

Forming their ratio, as in Eq. (3.7a), the length of $\gamma$ is bounded above by the integral

$$\int_{t_1}^{t_2} \left(\frac{g(r(t))}{g(\min_\gamma(r))}\right) \sqrt{\frac{r(t)}{\min_\gamma(r)}}\, dt. \quad (B8)$$

As $r(t) \geq \min_\gamma(r)$ in the integral ($r$ is a concave function of $t$ on $\alpha$ as well as on $\gamma$), and $g$ is non-decreasing, the term in parentheses in the integrand is no less than unity and acquires this minimal value when and only when $g$ is constant. Therefore, this ratio is minimal when $g$ is constant showing that the choice $f(r) = \sqrt{r}$, or a constant multiple thereof, is the best choice for this form of the argument.

However, there does exist a slightly more general argument. In writing Eq. (B4), we chose to write the right-hand side as a linear combination of $(1+\dot{r}^2)/r$ and $Q^2/r$, the idea being that we would then take advantage of the fact that $Q^2$ is non-negative and hence its appearance can be ignored if it appears with a negative coefficient. However, we can also include $m/r^2$ in this expression as well and then take advantage of its non-negativity in the same way. That is, we can write

$$\ddot{r} \leq -aQ^2/r - b(1+\dot{r}^2)/r - cm/r^2. \tag{B9}$$

for some $a$, $b$, and $c$. Comparing this to Eq. (B4), (being careful to note that these three quantities are not independent—they are related by Eq. (B3)) we find that

$$a + b = \frac{f'r}{f}, \tag{B10}$$

$$2b + c = 1. \tag{B11}$$

Therefore, if $a$, $b$, and $c$ are chosen so that they are all non-negative, we then have

$$\ddot{r} \leq -b(1+\dot{r}^2)/r, \tag{B12}$$

which is a slight generalization of Eq. (3.13). We can now get a slightly better bound than that established in Sec. III if we choose, for example, $f(r) = r^{1/3}$, $a = 0$, $b = 1/3$, and $c = 1/3$. Doing so, and carrying through the analysis, we find that we can establish theorem 1 with 9.31 in place of 10 (or the 9.7 mentioned and the end of Sec. III). As this improvement is so minor, we have not bothered with this messier generalization. Since, the choice $f = \sqrt{r}$, is not optimal, why not just take $f = 4\pi r^2$ which has a nice geometrical interpretation? We could, but then theorem 2 would have a number larger than 52 in place of 10. The choice $f = \sqrt{r}$ is a nice compromise between these two extremes.